\documentstyle[12pt,aaspp4,flushrt]{article}

\def\be{\begin{equation}}
\def\ee{\end{equation}}
\def\mathnew{\mathsurround=0pt}
\def\simov#1#2{\lower .5pt\vbox{\baselineskip0pt \lineskip-.5pt
        \ialign{$\mathnew#1\hfil##\hfil$\crcr#2\crcr\sim\crcr}}}
\def\simgreat{\mathrel{\mathpalette\simov >}}
\def\simless{\mathrel{\mathpalette\simov <}}

\def\yr{{\rm\,yr}}

\def\au{{\rm\,AU}}

\def\half{{\textstyle{1\over2}}}
\def\bfx{{\bf x}}
\def\bfy{{\bf y}}
\def\bfr{{\bf r}}
\def\bfJ{{\bf J}}
\def\bfw{{\bf w}}
\def\p{\partial}
\def\ffrac#1#2{{\textstyle\frac{#1}{#2}}}
\def\omit#1{}

\begin{document}

\title{The dynamics of Plutinos}

\author{Qingjuan Yu and Scott Tremaine}

\affil{Princeton University Observatory, Peyton Hall, \\
Princeton, NJ~08544-1001, USA}

\begin{abstract}
Plutinos are Kuiper-belt objects that share the 3:2 Neptune resonance with
Pluto. The long-term stability of Plutino orbits depends on their
eccentricity. Plutinos with eccentricities close to Pluto (fractional
eccentricity difference $\Delta e/e_p=|e-e_p|/e_p\simless 0.1$) can be stable
because the longitude difference librates, in a manner similar to the tadpole
and horseshoe libration in coorbital satellites.  Plutinos with $\Delta
e/e_p\simgreat 0.3$ can also be stable; the longitude difference circulates
and close encounters are possible, but the effects of Pluto are weak because
the encounter velocity is high. Orbits with intermediate eccentricity
differences are likely to be unstable over the age of the solar system, in the
sense that encounters with Pluto drive them out of the 3:2 Neptune resonance
and thus into close encounters with Neptune. This mechanism may be a source of
Jupiter-family comets. 
\end{abstract}


\section{Introduction}

The orbit of Pluto has a number of unusual features. It has the highest
eccentricity ($e_p=0.253$) and inclination ($i_p=17.1^\circ$) of any planet in
the solar system. It crosses Neptune's orbit and hence is susceptible to
strong perturbations during close encounters with that planet. However, close
encounters do not occur because Pluto is locked into a 3:2 orbital resonance
with Neptune, which ensures that conjunctions occur near Pluto's aphelion
(\cite{CH65}). More precisely, the critical argument
$3\lambda_p-2\lambda_n-\varpi_p$ librates around $180^\circ$ with a period of
$1.99\times 10^4$ yr and an amplitude of $82^\circ$; here $\lambda_p$ and
$\lambda_n$ are the mean longitudes of Pluto and Neptune and $\varpi_p$ is
Pluto's longitude of perihelion. Other resonances are present with longer
periods: for example, Pluto's argument of perihelion librates with an
amplitude of $23^\circ$ and a period of 3.8 Myr. In part because of its rich
set of resonances, Pluto's orbit is chaotic, although it exhibits no
large-scale irregular behavior over Gyr timescales (see \cite{MW97} for a
comprehensive review of Pluto's orbit). For reference, Pluto's 
semimajor axis and orbital period are $a_p=39.774\au$ and $P_p=250.85$ yr.

The most compelling explanation for Pluto's remarkable orbit was given by
\nocite{M93,M95} Malhotra (1993, 1995). Malhotra argues that Pluto formed in a
low-eccentricity, low-inclination orbit in the protoplanetary disk beyond
Neptune. Subsequent gravitational scattering and ejection of planetesimals in
the disk by all four giant planets caused Neptune's orbit to migrate outwards
(\cite{FI84}). As its orbit expands, Neptune's orbital resonances sweep
through the disk, first capturing Pluto into the 3:2 resonance and then
pumping up its eccentricity. If Pluto's orbit was circular before capture, its
present eccentricity implies that it was captured when Neptune's semimajor
axis was 0.814 times its current value or $24.6\au$ (eq. \ref{eq:renu}). This
process may also excite Pluto's inclination although the details are less
certain (\cite{M98}).

Malhotra's argument predicts that most Kuiper belt objects with $30\au\simless
a\simless 50\au$ should also be captured---and presently located---in Neptune
resonances (\cite{M95}). This prediction has proved to be correct: of the
$\sim 90$ Kuiper belt objects with reliable orbits as of 1999 January 1, over
30\% have semimajor axes within 1\% of the 3:2 resonance (although this number
is exaggerated by observational selection effects). These objects have come to
be called Plutinos, since they share the 3:2 resonance with Pluto (see
\cite{MDL99} for a recent review of the Kuiper belt).

Almost all studies of the dynamics of the Kuiper belt so far have neglected
the gravitational influence of Pluto, because of its small mass
($M_p/M_\odot=7.40\times 10^{-9}$ for the Pluto-Charon system,
\cite{S92,TW97}). However, like the Trojan asteroids and Jupiter, or the Saturn
coorbital satellites Janus and Epimetheus, Pluto and the Plutinos share a
common semimajor axis and hence even the weak gravitational force from Pluto
can have a substantial influence on the longitude of a Plutino relative to
Pluto. A crude illustration of the importance of Pluto's gravity is to note
that the half-width of the 3:2 resonance, $(\Delta a/a)_{\rm res}\simeq 0.01$
for $0.2\simless e\simless 0.3$ (the maximum fractional amplitude of stable
libration in semi-major axis, \cite{M96}), is only a few times larger than the
Hill radius of Pluto, $(\Delta a/a)_{\rm H}=(M_p/3M_\odot)^{1/3}=0.0014$.

The goal of this paper is to explore the dynamical interactions between Pluto
and Plutinos and their consequences for the present structure of the Kuiper
belt. Section 2 provides an approximate analytical description of the
interactions,  \S 3 describes the results of numerical orbit
integrations, and \S 4 contains a discussion.

\section{Analysis}

We examine a simplified model solar system containing only the Sun and
Neptune, with masses $M_\odot$ and $M_n$; we assume that Neptune's orbit is
circular and neglect all orbital inclinations. We describe the motion of the
Plutino using the canonical variables
\begin{eqnarray}
x_1=(GM_\odot a)^{1/2},\qquad\qquad & & y_1=\lambda, \nonumber \\
x_2=(GM_\odot a)^{1/2}[1-(1-e^2)^{1/2}],\qquad & &
y_2=-\varpi,
\label{eq:canvar}
\end{eqnarray}
where $a$ and $e$ are the semi-major axis and eccentricity, $\lambda$ and
$\varpi$ are the mean longitude and longitude of perihelion. The same
variables for Neptune or Pluto are denoted by adding a subscript ``n'' or
``p''.

We consider only the resonant perturbations exerted by
Neptune, which can only depend on angles as an even function of the combination
$3y_1-2y_{n1}+y_2$. Thus the Hamiltonian of a Plutino may be written 
\be 
H_0(\bfx,\bfy,t)=H_K(x_1)+A(\bfx,3y_1-2y_{n1}+y_2)+B(\bfx,\bfy,t), 
\ee 
where $H_K(x_1)=-\half(GM_\odot)^2/x_1^2$ is the Kepler Hamiltonian, $A$ is
the resonant potential from Neptune, and $B(\bfx,\bfy,t)$ is the
potential from Pluto. The same Hamiltonian describes the motion of Pluto if
we set $B=0$.

Now impose a canonical transformation to new variables $(\bfJ,\bfw)$ defined
by the generating function 
\be
S(\bfJ,\bfy,t)=J_1(3y_1-2y_{n1}+y_2)+\ffrac{1}{3}J_2(y_2-y_{p2}).  
\ee
Thus $w_1=\partial S/\partial J_1=3y_1-2y_{n1}+y_2$, 
$w_2=\partial S/\partial J_2=\frac{1}{3}(y_2-y_{p2})$, while
$x_1=\partial S/\partial y_1=3J_1$, 
$x_2=\partial S/\partial y_2=J_1+\frac{1}{3}J_2$, and the new Hamiltonian is 
\be 
H(\bfJ,\bfw,t)=H_0+{\p S\over \p t}= H_K(3J_1)-2\dot y_{n1}J_1-\ffrac{1}{3}\dot
y_{p2}J_2+A(3J_1,J_1+\ffrac{1}{3}J_2,w_1)+B,
\label{eq:ham}
\ee 
where $\dot y_{n1}$ is the mean motion of Neptune and $-\dot y_{p2}$  is the
apsidal precession rate of Pluto.

To proceed further we make use of the fact that $B/A=\hbox{O}(M_p/M_n)\ll
1$. Thus we can divide the motion of a Plutino into ``fast'' and ``slow''
parts. The fast motion is determined by the Kepler Hamiltonian $H_K$ and the
resonant potential $A$ (this is the opposite of normal usage, where the
resonant perturbations from Neptune are regarded as ``slow'' compared to
non-resonant perturbations). The slow variations are caused by the Pluto
potential $B$.

First we examine the fast motion. We drop the potential $B$, so that $w_2$ is
ignorable and $J_2=3x_2-x_1=(GM_\odot a)^{1/2}[2-3(1-e^2)^{1/2}]$ is a
constant of the motion. Thus if Pluto was captured into resonance from a
circular orbit with semimajor axis $a_i$, its present semimajor axis and
eccentricity are related by 
\be
e^2=\ffrac{5}{9}-\ffrac{4}{9}(a_i/a)^{1/2}-\ffrac{1}{9}(a_i/a).
\label{eq:renu}
\ee

We write $J_1=J_{1r}+\Delta
J_1$ where $J_{1r}$ is chosen to satisfy the resonance condition for the
Kepler Hamiltonian, 
\be
2\dot y_{n1}=3\dot y_1=3 {dH_K\over dx_1}\qquad\hbox{or} \qquad
J_{1r}=\ffrac{1}{3}x_{1r}={(GM_\odot)^{2/3}\over (18\dot y_{n1})^{1/3}}.
\label{eq:chosen}
\ee
Since $A/H_K=\hbox{O}(M_n/M_\odot)\ll1$ we expect $|\Delta J_1|\ll J_{1r}$, so
we can expand $H_K$ to second order in $\Delta J_1$; dropping unimportant
constant terms the fast motion is determined by the Hamiltonian 
\begin{eqnarray} 
H_f(\Delta J_1,w_1) & = & \frac{9}{2}\left({d^2H_K\over
dx_1^2}\right)_{x_{1r}} (\Delta J_1)^2+A(3J_{1r}+3\Delta J_1,J_{1r}+\Delta
J_1+\ffrac{1}{3}J_2,w_1) \nonumber \\ & = & -{27\over 2a^2}(\Delta
J_1)^2+A(3J_{1r}+3\Delta J_1,J_{1r}+\Delta J_1+\ffrac{1}{3}J_2,w_1).
\label{eq:pend}
\end{eqnarray} 
The Hamiltonian is autonomous and hence has a conserved energy
$E_f=H_f$ and action $I=(2\pi)^{-1}\oint \Delta J_1 dw_1$. The motion 
is along the level surfaces of $H_f$ in the $(\Delta J_1,w_1)$ plane and
typically consists of either libration ($w_1$ oscillates between fixed limits)
or circulation ($w_1$ increases or decreases without reversing), just as in
the case of the pendulum Hamiltonian. The stable equilibrium solutions
(i.e. zero-amplitude libration) are given by
\be
\Delta J_1={a^2\over 9}\left({\p A\over\p x_1}+\frac{1}{3}{\p A\over\p
x_2}\right), \qquad {\p A\over \p w_1}=0, \qquad {\p^2A\over \p w_1^2}<0.
\label{eq:eq}
\ee 
The slow motion is determined by averaging the Hamiltonian (\ref{eq:ham}) over
the fast motion: 
\be 
H_s(J_2,w_2,t)=E_f(J_2)-\ffrac{1}{3}\dot y_{p2}J_2+\langle B\rangle.
\label{eq:hamslow}  
\ee 
Here $\langle\cdot\rangle$ indicates an
average over one period of the fast motion. The fast energy $E_f$ depends
on $J_2$ through the constraint that the fast action $I$ is adiabatically
invariant.

\subsection{Solutions with zero-amplitude libration}

The solutions to the fast and slow equations of motion are particularly simple
in the case where the fast libration amplitude is zero for both Pluto and the
Plutino. This approximation is not particularly realistic---the 
libration amplitude of Pluto is $82^\circ$---but illustrates the principal
features of the Plutino motions. 

In this case the fast energy is 
\be
E_f=A(3J_{1r},J_{1r}+\ffrac{1}{3}J_2,w_{1r}) 
\ee
where $w_{1r}$ is the equilibrium angle given by equation (\ref{eq:eq}) and we
have dropped much smaller terms that are O$(A^2)$. For simplicity we shall
assume that there is only one stable equilibrium point, that is, one solution
to equations (\ref{eq:eq}) for given $J_2$. At the equilibrium point the fast
action is $I=0$, and the slow Hamiltonian (\ref{eq:hamslow}) is 
\be
H_s(J_2,w_2,t)=A(3J_{1r},J_{1r}+\ffrac{1}{3}J_2,w_{1r}) -\ffrac{1}{3}\dot
y_{p2}J_2 + \langle
B(3J_{1r},J_{1r}+\ffrac{1}{3}J_2,y_1,y_2,t)\rangle+\hbox{O}(A^2), 
\ee 
where $y_1=\frac{1}{3}(w_{1r}+2y_{n1}-y_2)$. Since Pluto also is assumed to
have zero libration amplitude, $y_{p1}=\frac{1}{3}(w_{1r}+2y_{n1}-y_{p2})$. 
Thus 
\be
3(y_{p1}-y_1)=y_2-y_{p2}=3w_2 \qquad \hbox{mod}\,(2\pi);
\label{eq:resrel}
\ee
that is, the difference in longitude of perihelion between Pluto and the
Plutino is three times the difference in mean longitude. The same result will
hold true on average even if the libration amplitudes are non-zero. 

Now let us assume in addition that the eccentricities of both Pluto and the
Plutino are small. Since their semi-major axes are the same, the gravitational
potential from Pluto at the Plutino may be written
\be 
B=GM_p\left({\bfr\cdot\bfr_p\over|\bfr_p|^3}-{1\over|\bfr-\bfr_p|}\right) = 
{GM_p\over a}\left[\cos(y_1-y_{p1})-{1\over 2|\sin\half(y_1-y_{p1})|} 
\right].
\ee
Using equation (\ref{eq:resrel}) this simplifies to
\be 
B={GM_p\over a}\left(\cos w_2-{1\over 2|\sin\half w_2|} \right).
\ee
Moreover 
\be
{dy_{p2}\over dt}={\p A\over \p x_2}(3J_{1r},J_{1r}+\ffrac{1}{3}J_{p2},w_{1r}) 
+\hbox{O}(A^2);
\ee
thus the slow Hamiltonian can be rewritten as
\begin{eqnarray}
H_s(J_2,w_2)=A(3J_{1r},J_{1r}+\ffrac{1}{3}J_2,w_{1r}) 
-\ffrac{1}{3}J_2{\p A\over\p x_2}(3J_{1r},J_{1r}+\ffrac{1}{3}J_{p2},w_{1r}) 
\nonumber \\
+{GM_p\over a}\left(\cos w_2-{1\over 2|\sin\half w_2|} \right)
+\hbox{O}(A^2).
\end{eqnarray}
The interesting behaviour occurs when the actions $J_2$ and $J_{p2}$ are
similar (i.e. the eccentricities of Pluto and the Plutino are similar), so we
write $J_2=J_{p2}+\Delta J_2$ and expand $A$ to second order in $\Delta
J_2$. Dropping unimportant constants and terms of O$(A^2)$ we have
\be
H_s(\Delta J_2,w_2)=\frac{1}{18}\Delta J_2^2A_{22}
(3J_{1r},J_{1r}+\ffrac{1}{3}J_{p2},w_{1r}) 
+{GM_p\over a}\left(\cos w_2-{1\over 2|\sin\half w_2|} \right).
\label{eq:tro}
\ee
where $A_{22}=\p^2A/\p x_2^2$. 

This Hamiltonian is strongly reminiscent of the Hamiltonian for a test
particle coorbiting with a satellite, 
\be
H_c(\Delta x,w)=-{3\over 2a^2}\Delta x^2 +{GM\over a}\left(\cos
w-{1\over 2|\sin\half w|} \right);
\label{eq:hamc}
\ee 
here $\Delta x=x_1-x_{s1}$ and $w=\lambda_1-\lambda_{s1}$ are conjugate
variables, and the subscript $s$ denotes orbital elements of the satellite.
In this case the torques from the satellite lead to changes in  semi-major
axis; for a Plutino the semi-major axis is locked to Neptune's by the
resonance, so torques from Pluto lead to changes in the eccentricity instead. 

When $A_{22}<0$ many of the features of orbits in the slow Hamiltonian
(\ref{eq:tro}) follow immediately from the analogy with the Hamiltonian
(\ref{eq:hamc}), which has been studied by many authors (e.g. Yoder et
al. 1983, Namouni et al. 1999).  The trajectories are determined by the level
surfaces of the Hamiltonian. The equilibrium solutions correspond to the
triangular Lagrange points in the coorbital case: $\Delta J_2=0$, $w_2=\pm
60^\circ$, $H_s=-\half GM_p/a$; the eccentricity of the Plutino equals the
eccentricity of Pluto, the mean longitude leads or lags by $60^\circ$, and the
perihelia are $180^\circ$ apart. These solutions are maxima of the potential
from Pluto. For smaller values of $H_s$, the orbits librate around the
triangular points (``tadpole orbits''). Small-amplitude tadpole librations
have frequency $\omega$ given by \be \omega^2=-\frac{1}{4}A_{22}{GM_p\over a}.
\label{eq:freq}
\ee
The tadpole orbits merge at the separatrix orbit,
$H_s=-\frac{3}{2}GM_p/a$; for this orbit the minimum separation is
$w_{2,\rm min}=23.91^\circ$. Even smaller values of $H_s$ yield ``horseshoe''
orbits, with turning points at $\pm w_{2,\rm min}$ where $H_s=(GM_p/a)(\cos
w_{2,\rm min}-\half|\sin \half w_{2,\rm min}|^{-1})$. For all tadpole and
horseshoe orbits, the maximum and minimum values of $\Delta J_2$ occur at
$w_2=\pm 60^\circ$, and are given by
\be
\Delta J_2=\pm\left[{18\over A_{22}}\left(H_s+{GM_p\over
2a}\right)\right]^{1/2}. 
\label{eq:sep}
\ee
Eventually the theory breaks down, when $w_{2,\rm min}$ is small enough that
adiabatic invariance is no longer a valid approximation.

\subsection{The resonant potential from Neptune}

For quantitative applications we must evaluate the resonant Neptune potential
$A(\bfx,w_1)$. For small eccentricities, the potential can be derived
analytically, 
\begin{eqnarray}
A(\bfx,w_1) &= & -{GM_n\over a}\bigg[\half b_{1/2}^{(0)}(\alpha)
+\ffrac{1}{8}e^2(2\alpha D+\alpha^2D^2)b_{1/2}^{(0)}(\alpha) +
\half e(5+\alpha D)b_{1/2}^{(2)}(\alpha)\cos w_1  \nonumber \\
& & +\ffrac{1}{8}e^2(104 +22\alpha D+\alpha^2D^2)b_{1/2}^{(4)}(\alpha)
\cos 2w_1
+\hbox{O}(e^3)\bigg],
\label{eq:anal} 
\end{eqnarray}
where $\alpha=a_n/a<1$, $D=d/d\alpha$,
\be
b_{1/2}^{(j)}(\alpha)={1\over\pi}\int_0^{2\pi} {d\phi \cos j\phi
\over (1-2\alpha\cos\phi+\alpha^2)^{1/2}}\ee
is a Laplace coefficient, Neptune is assumed to be on a circular orbit, and
inclinations are neglected.

Unfortunately, in the present case the high eccentricity of the Plutino orbits
makes this expansion invalid. However, we may determine $A(\bfx,w_1)$
numerically for given actions $\bfx$ by averaging the gravitational potential
from Neptune over $y_2$ at fixed $w_1$, that is, 
\be
A(\bfx,w_1)=-{GM_n\over a}F(\bfx,w_1),
\label{eq:resdef}
\ee
where 
\be
F(\bfx,w_1)=
{a\over 6\pi}\int_0^{6\pi} dy_2\left ({1\over
|\bfr_n-\bfr|}-{\bfr\cdot\bfr_n\over |\bfr_n|^3}\right)_{\bfx,w_1,y_2}+
\hbox{constant};
\label{eq:AF}
\ee the unimportant constant is chosen so that $A=0$ for circular orbits, i.e.
$F(x_1,0,w_1)=0$.  It can be shown analytically that the contribution from the
second (indirect) term in the integrand vanishes.

We shall also write 
\be
A_{22}={\p^2 A\over \p x_2^2}\equiv -{M_n\over M_\odot a^2}
F_{22}(\bfx,w_1)\quad
\hbox{where} \quad F_{22}(\bfx,w_1)=x_1^2{\p^2F(\bfx,w_1)\over \p x_2^2}.
\label{eq:twotwo} 
\ee

Figure \ref{fig:contone}\ plots the contours of $F(\bfx,w_1)$ at the resonant
semimajor axis $x_{1r}$, as obtained from equation (\ref{eq:AF}). The
potential is singular for collision orbits, which for small eccentricity
satisfy 
\be 
\cos w_1={a-a_n\over ea}.
\label{eq:coll} 
\ee The conditions (\ref{eq:eq}) for stable zero-amplitude libration are
satisfied if and only if $w_1=w_{1r}=\pi$ or $w_1=0$ and
$e>1-a_n/a=0.237$. Figure \ref{fig:corot180}\ and \ref{fig:corot0}\ show
examples of these two solutions, plotted in a reference frame corotating with
Neptune. Orbits of the first kind are similar to Pluto's, although with
smaller libration amplitude (compare Fig. 4 of \cite{MW97}). Orbits of the
second kind (Fig. \ref{fig:corot0}) were discussed by \nocite{M96} Malhotra
(1996), who calls them ``perihelion librators''. We shall not discuss these
further, since they do not appear to form naturally during resonance capture
of initially circular orbits; moreover for $e\simgreat 0.35$ they are likely
to be unstable, since they cross Uranus's orbit and thus are subject to close
encounters and collisions with that planet.

Figure \ref{fig:approx}\ plots $F(x_{1r},x_2,\pi)$ and
$F_{22}(x_{1r},x_2,\pi)$; at the eccentricity of Pluto, corresponding to
$x_2/(GM_\odot a)^{1/2}=0.0325$, we have $F(x_{1r},x_2,\pi)=-0.313$ and
$F_{22}(x_{1r},x_2,\pi)=79.3$.  Thus, for example, the eccentricity
oscillation in the separatrix orbit that marks the boundary between tadpole
and horseshoe orbits has amplitude $\Delta e =0.007$ (eq. \ref{eq:sep}) and
the period of libration of small tadpole orbits is $2\pi/\omega=9.1\times
10^{7}$ yr (eq. \ref{eq:freq}).

For our purposes it is sufficient to work with the following numerical
approximation to the resonant potential:
\be
\widetilde F(\bfx,w_1)=-\frac{0.584+0.130e}{1+1.709e}\ln
\left|1-4.222e\cos w_1\right|,
\label{eq:F}
\ee
where $x_2=x_{1r}[1-(1-e^2)^{1/2}]$. This approximation is chosen to match the
resonant potential at the resonant semimajor axis $x_1=x_{1r}$; the dependence
on the relative semimajor axes of Neptune and the Plutino is suppressed since
the effects of this potential are only important near resonance. The
logarithmic factor is chosen to reproduce the singularity in the resonant
potential near the collision orbits defined approximately by equation
(\ref{eq:coll}). The approximation formula also matches the analytic formula
(\ref{eq:anal}) to O$(e^2)$ at $w_1=\pi$. 

Figure \ref{fig:conttwo}\ shows the contour plot analogous to Figure
\ref{fig:conttwo}\ for the approximate resonant potential $\widetilde F$, and
the triangles in Figure \ref{fig:approx}\ show $\widetilde F$ and $\widetilde
F_{22}$. The agreement is very good, especially considering that errors are
amplified by taking the two derivatives required to generate $\widetilde
F_{22}$.

\section{Numerical experiments}

We follow the orbital evolution of Pluto and a Plutino in a simplified version
of the Sun-Neptune-Pluto-Plutino four-body system that isolates the resonant
potential from Neptune.  Neptune is assumed to have a circular orbit that
migrates outward according to the rule (\cite{M93}) \be a_n(t)=a_f-\Delta
a\exp(-t/\tau), \ee where $a_f=30.17\au$ is Neptune's present semimajor axis,
$\Delta a=6\au$, and $\tau=1.5$ Myr. Thus Neptune's initial semimajor axis is
$24.17\au$ and the initial location of the 3:2 orbital resonance is
$31.67\au$.

The initial eccentricity of Pluto is taken to be zero and its initial
semimajor axis is $33\au$, as required so that its present eccentricity
matches the observed value (eq. \ref{eq:renu}). We followed 160 test
particles, with initial semimajor axes distributed uniformly in the range
$[31\au,39\au]$ and eccentricities distributed uniformly in the range
$[0,0.03]$. The inclinations of Pluto and the test particles are chosen
randomly in the range $[0,3^\circ]$ and their angular elements are chosen
randomly from $[0,2\pi]$.  Pluto and the test particles feel the resonant
potential from Neptune, as defined by equations (\ref{eq:resdef}) and
(\ref{eq:F}), but no other Neptune forces. The effects of the resonant Neptune
potential on the orbital elements of Pluto and the test particles are followed
using Lagrange's equations.

The test particles do not influence Pluto or one
another. However, they are subject to the gravitational potential from
Pluto, 
\be 
B(\bfx,\bfy,t)=-GM_p\left({1\over
|\bfr-\bfr_p|}-{\bfr\cdot\bfr_p\over|\bfr_p|^3}\right);
\label{eq:B}
\ee
the effects of this potential on the orbital elements of the test particles are
followed using Gauss's equations.

The evolution of Pluto and the test particles is followed for 0.45 Gyr or 10\%
of the age of the solar system.

\section{Results}

Of the 160 test particles, all but 12 are captured into the 3:2 resonance with
Neptune, in the sense that their final semimajor axes are close to
$(3/2)^{2/3}a_n$ and their eccentricities are near the
prediction of equation (\ref{eq:renu}), as shown in Figure \ref{fig:renu}.
The 12 particles that are not captured lie inside the location of Neptune's
3:2 resonance at the start of the calculation, $(3/2)^{2/3}\times
24.17\au=31.67\au$, and would presumably be captured into other resonances if
we used the full Neptune potential to work out their motion. We have verified
this presumption by conducting shorter integrations ($1\times10^7 \yr$) using
the same initial conditions but the complete Neptune potential.  In this case
all but 15 of the 160 particles were captured into the 3:2 resonance; the
remainder were captured into the 4:3, 5:3 or 7:5 resonances.

The behavior of the test particles in the 3:2 resonance (henceforth Plutinos)
falls into the following broad classes:

\begin{itemize}

\item Tadpole orbits (5 particles): these have longitude difference
$y_1-y_{p1}$ and differences in longitude of perihelion $y_{p2}-y_2$ that
librate around the leading or trailing Lagrange point of Pluto (Figure
\ref{fig:tadpole}). 
(Note that the libration center for the orbit in this
Figure is $y_1-y_{p1}\simeq 100^\circ$, not $60^\circ$ as implied by the
analysis in \S 2.1. This discrepancy arises because Pluto has a high
eccentricity, while our analysis is only valid for near-circular
orbits. Similarly, the perihelion difference librates around
$\varpi-\varpi_p\simeq 300^\circ$, three times the difference in mean
longitude as required by eq. \ref{eq:resrel}.) The
tadpoles show no evidence of chaotic behavior or secular evolution over the
length of our integration. The analysis in \S 2.2 suggests that the maximum
eccentricity difference for these orbits is $\Delta e\simeq 0.007$; this
requires in turn that their initial semimajor axes must have been close to
Pluto's, as is seen to be the case in Figure \ref{fig:renu}.

\item Horseshoe orbits (19 particles): the longitude difference oscillates
around $180^\circ$, with jumps in the Plutino eccentricity at the extrema of
the longitude oscillation, as predicted by the analysis of \S 2.1 (Figures
\ref{fig:horseshoe1}, \ref{fig:horseshoe2}).  The motion appears stable over
the length of our integration although there are significant variations in
semimajor axis oscillations during the course of the integration, and some
horseshoes may evolve into transitional orbits over longer time intervals.

\item Transitional orbits (43 particles): these show irregular behavior or
transitions between libration and circulation of the longitude difference
(Figure \ref{fig:trans}). When the longitude difference circulates, the
particles are no longer protected from close encounters with Pluto. However,
the particles remain in the 3:2 Neptune resonance in the sense that the
resonant angle $w_1$ continues to librate.

\item Doubly transitional orbits (2 particles): Like transitional orbits,
these show libration-circulation transitions in the longitude difference, but
in addition they show irregular behavior in the resonant angle $w_1$, leading
eventually to a transition of $w_1$ from libration to circulation (Figure
\ref{fig:dtrans}). Although only 2 particles in our simulation exhibit this
behavior, a number of others show growing amplitude in the $w_1$ libration and
will probably move into this class in less than the age of the solar
system. Such orbits are normally short-lived since once $w_1$ circulates, they
are no longer protected from close encounters with Neptune.

\item Irregular circulating orbits (17 particles): the longitude difference
circulates throughout the integration. Pluto induces irregular behavior
(Figure \ref{fig:unaffect2}), but the Neptune resonance is preserved in the
sense that $w_1$ continues to librate, at least over the span of our
integration.

\item Regular circulating orbits (62 particles): the longitude difference
circulates throughout the integration, but the orbits appear
fairly regular (Figure \ref{fig:unaffect1}). Generally, the orbits with
larger eccentricity differences are more regular, because the
encounter velocity with Pluto is higher so the perturbations from close
encounters are smaller. 

\end{itemize}

These classes represent a sequence in eccentricity difference: the
typical eccentricity difference $|e-e_p|$ is smallest for tadpoles and largest
for orbits unaffected by Pluto. Because the Plutino eccentricity is determined
by the semimajor axis at the time of resonant capture (eq. \ref{eq:renu}), the
classes also reflect the initial semimajor axes of the Plutinos: the tadpoles
and horseshoes all have initial semimajor axes in the range 32.2\au--34.2\au\ 
(i.e. close to Pluto's initial semimajor axis of 33\au). The transitional
and irregular circulating Plutinos mostly have initial semimajor axes
in the range 31.7\au--36\au, and the regular circulating Plutinos have initial
semimajor axes concentrated in the range 35\au--39\au.

\section{Discussion}

Test particles captured into the 3:2 Neptune resonance (Plutinos) have a
complex range of dynamical interactions with Pluto. The strength of the
interaction depends on the difference in eccentricity between the test
particle and Pluto, and thus on the difference in initial semi-major axis if
the initial orbits were circular and capture occurred through outward
migration of Neptune. Plutinos are stable only if the eccentricity difference
$\Delta e$ is small ($\Delta e\simless 0.02$ from Figure \ref{fig:renu}), in
which case the Plutinos librate on tadpole or horseshoe orbits; or if the
eccentricity difference is large ($\Delta e\simgreat 0.06$), in which case the
longitude difference circulates but relative velocity at encounter is high
enough that Pluto has little effect. Unstable orbits at intermediate $\Delta
e$ can be driven out of the 3:2 Neptune resonance by interactions with Pluto,
and thereafter are short-lived because of close encounters with Neptune. Thus
we expect that the population of Plutinos has decayed over time, although
determining the survival fraction will require integrations over the lifetime
of the solar system using the full Neptune potential.

The long-term behavior of orbits in the 3:2 Neptune resonance is central to the
origin of Jupiter-family comets. The usual explanation is that slow chaotic
diffusion and collisional kicks drive Plutinos out of the 3:2 resonance, after
which they are subjected to close encounters with the giant planets and
eventually evolve into Jupiter-family comets (\cite{mo97}). Our results suggest
that Pluto-induced evolution of Plutinos onto Neptune-crossing orbits may
contribute to or even dominate the flux of Jupiter-family comets.

Our results also enhance the motivation to obtain accurate orbits for
Kuiper-belt objects, and fuel speculation that the formation of the
Pluto-Charon binary may be linked to interactions between Pluto and Plutinos.

This research was supported in part by NASA Grant NAG5-7310. We thank Matt
Holman, Renu Malhotra, and Fathi Namouni for discussions and advice.

\newpage

\figcaption[fig1.gif]{A contour plot of the resonant gravitational potential
$F(\bfx,w_1)$ at the resonant semimajor axis, $x_1=x_{1r}$
(eq. \ref{eq:chosen}), as obtained from equation (\ref{eq:AF}). The ordinate
is eccentricity, which is a proxy for $x_2=x_{1r}[1-(1-e^2)^{1/2}]$. The
contour levels are uniformly spaced at intervals of 0.01; positive contours are
solid and negative contours are dotted. \label{fig:contone}}

\figcaption[fig2.gif]{The orbit of a Plutino for 20,000 years in a reference
frame corotating with Neptune. Initially the resonant angle $w_1=180^{\circ}$
and the eccentricity $e=0.39$. The high eccentricity was chosen to illustrate
the character of the orbit; in a more realistic integration this orbit would be
unstable because it crosses the orbit of Uranus. \label{fig:corot180}}

\figcaption[fig3.gif]{The orbit of a Plutino for 20,000 years in a reference
frame corotating with Neptune. Initially the resonant angle $w_1=0^{\circ}$ and
$e=0.49$. In a more realistic integration this orbit would be unstable, since
it crosses the orbit of Uranus. \label{fig:corot0}}

\figcaption[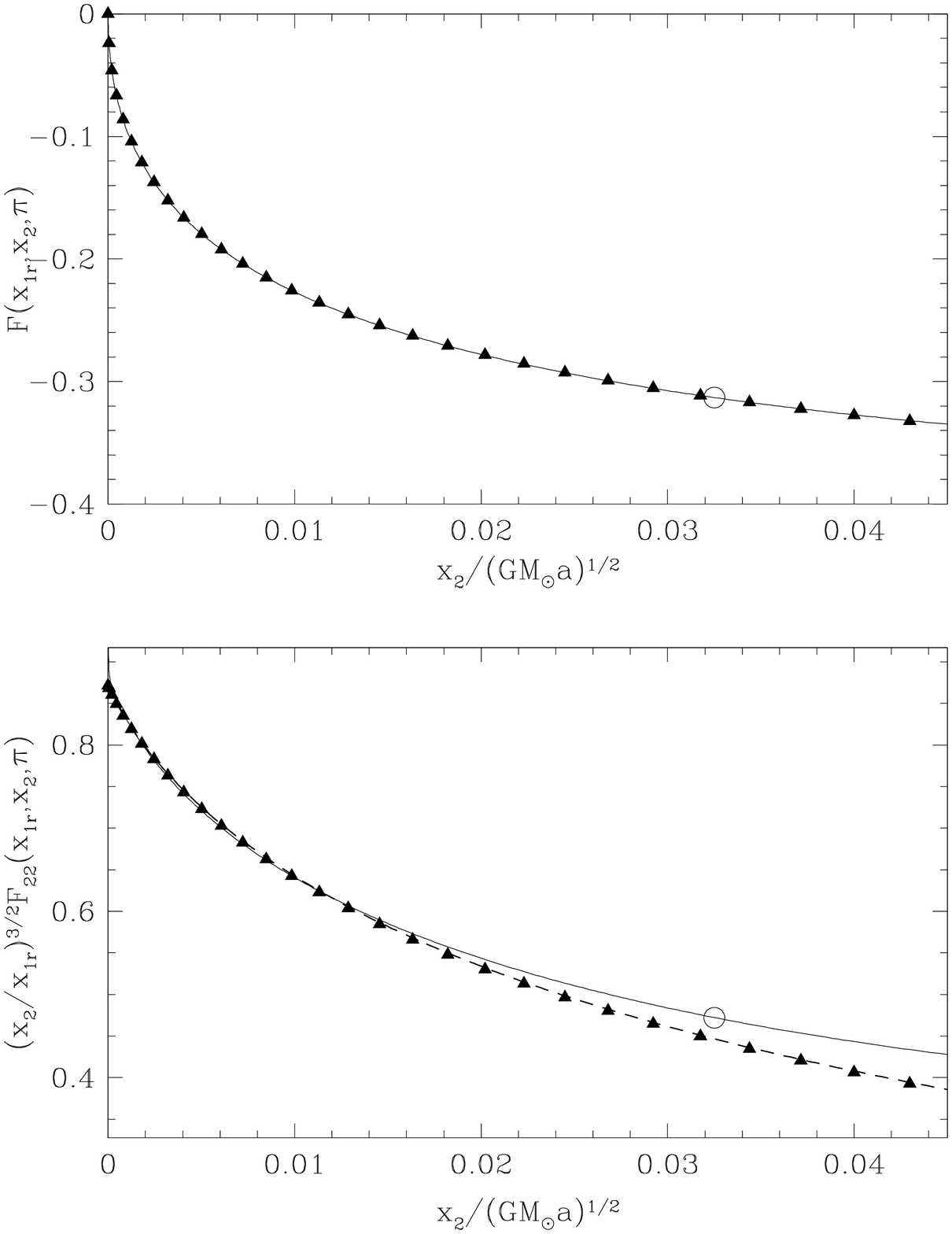]{The solid curves plot $F(x_{1r},x_2,\pi)$ and
$(x_2/x_{1r})^{3/2}F_{22}(x_{1r},x_2,\pi)$ as determined by the numerical
integral (\ref{eq:AF}) and equation (\ref{eq:twotwo}) (the multiplicative
factor in front of $F_{22}$ is used because $F_{22}\propto x_2^{3/2}$ as
$x_2\to0$). The solid triangles and dashed line show the approximations
$\widetilde F$ and $\widetilde F_{22}$ given by equation (\ref{eq:F}). For
reference, the eccentricity of Pluto, $e=0.253$, is marked by an open circle
at $x_2/(GM_\odot a)^{1/2}=0.0325$. \label{fig:approx}}

\figcaption[fig5.gif]{The contour graph of the approximation $\widetilde F$
(eq. \ref{eq:F}) to the resonant gravitational potential. The contour levels
are the same as in Figure \ref{fig:contone}. \label{fig:conttwo}}

\figcaption[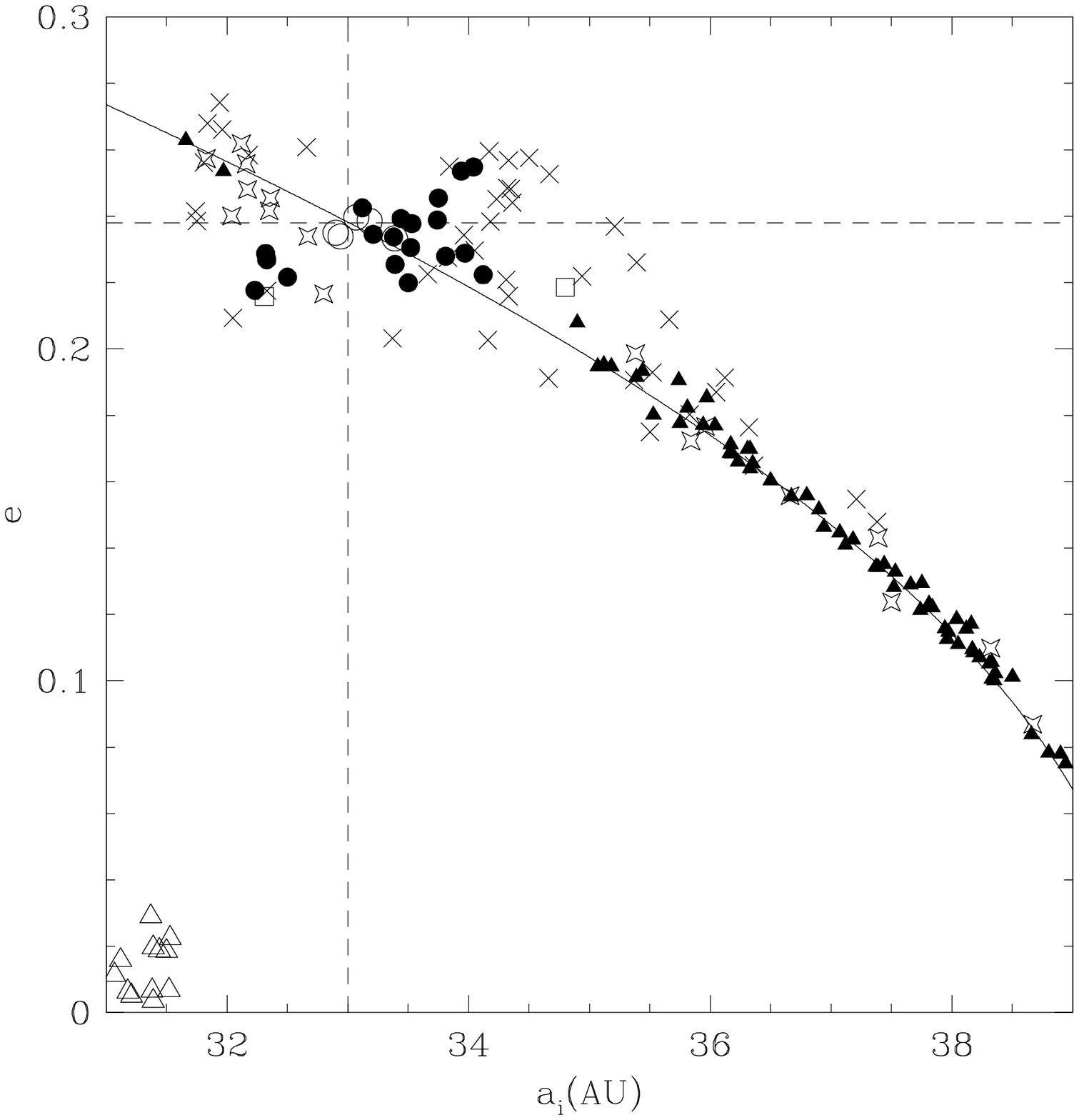]{The final eccentricity of the test particles as a
function of their initial semimajor axis. The solid curve is the prediction of
equation (\ref{eq:renu}), and the crossing point of the dashed lines shows the
location of Pluto, computed with the same equation. The symbols denote the orbit
classes discussed in \S 4: tadpole orbits (open circles), horseshoe orbits
(solid circles), transitional orbits (crosses), doubly transitional orbits
(open squares), irregular circulating orbits (starred squares), regular
circulating orbits (solid triangles), orbits not captured into the 3:2
resonance (open triangles). \label{fig:renu}}

\figcaption[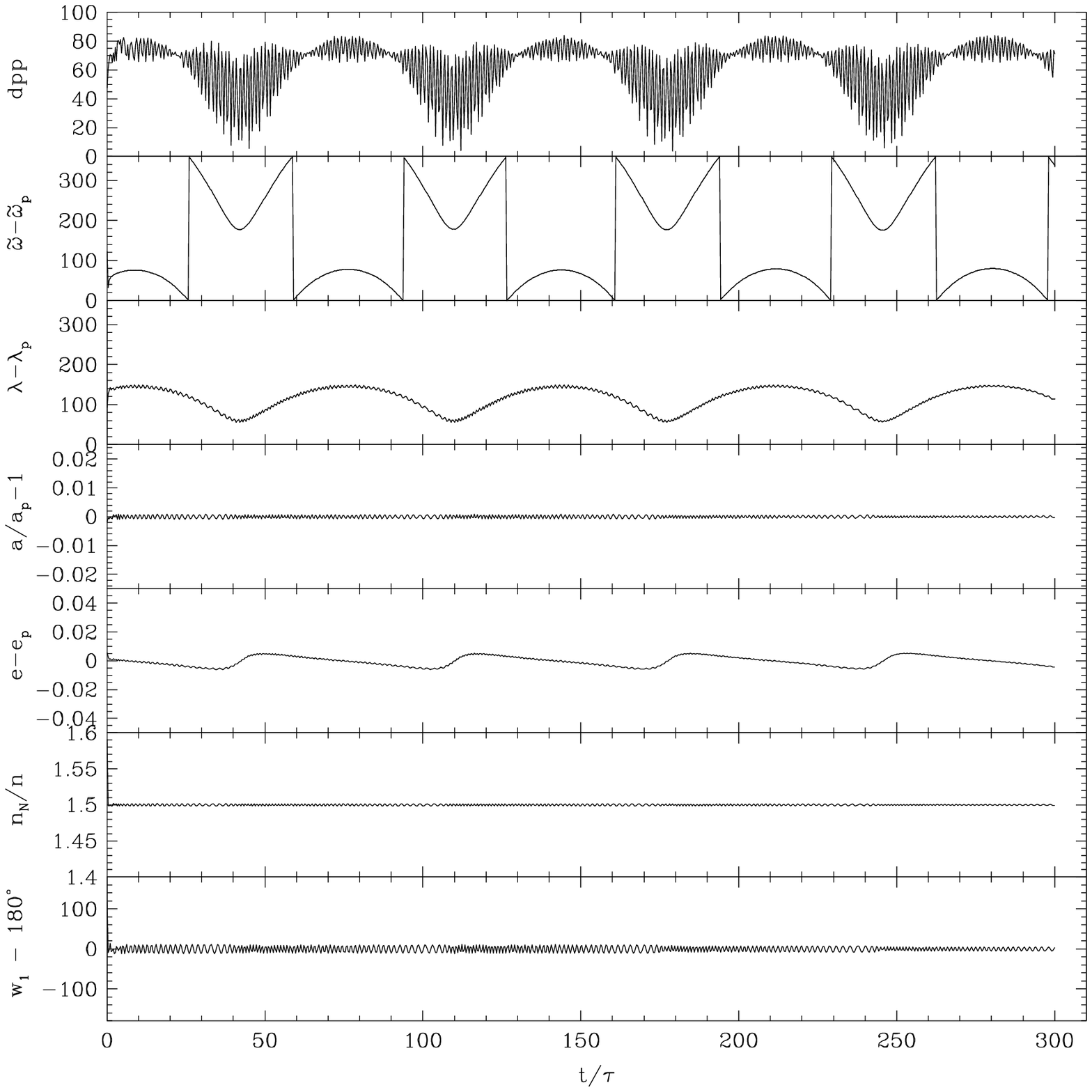]{A tadpole orbit.  In the first panel ``dpp'' denotes the
distance between Pluto and the Plutino in AU. Subsequent panels show longitude
of perihelion, mean longitude, semimajor axis, eccentricity, orbital period
ratio to Neptune, and the resonant angle of the Plutino; the detailed
definitions of the orbital elements are in the text. \label{fig:tadpole}}

\figcaption[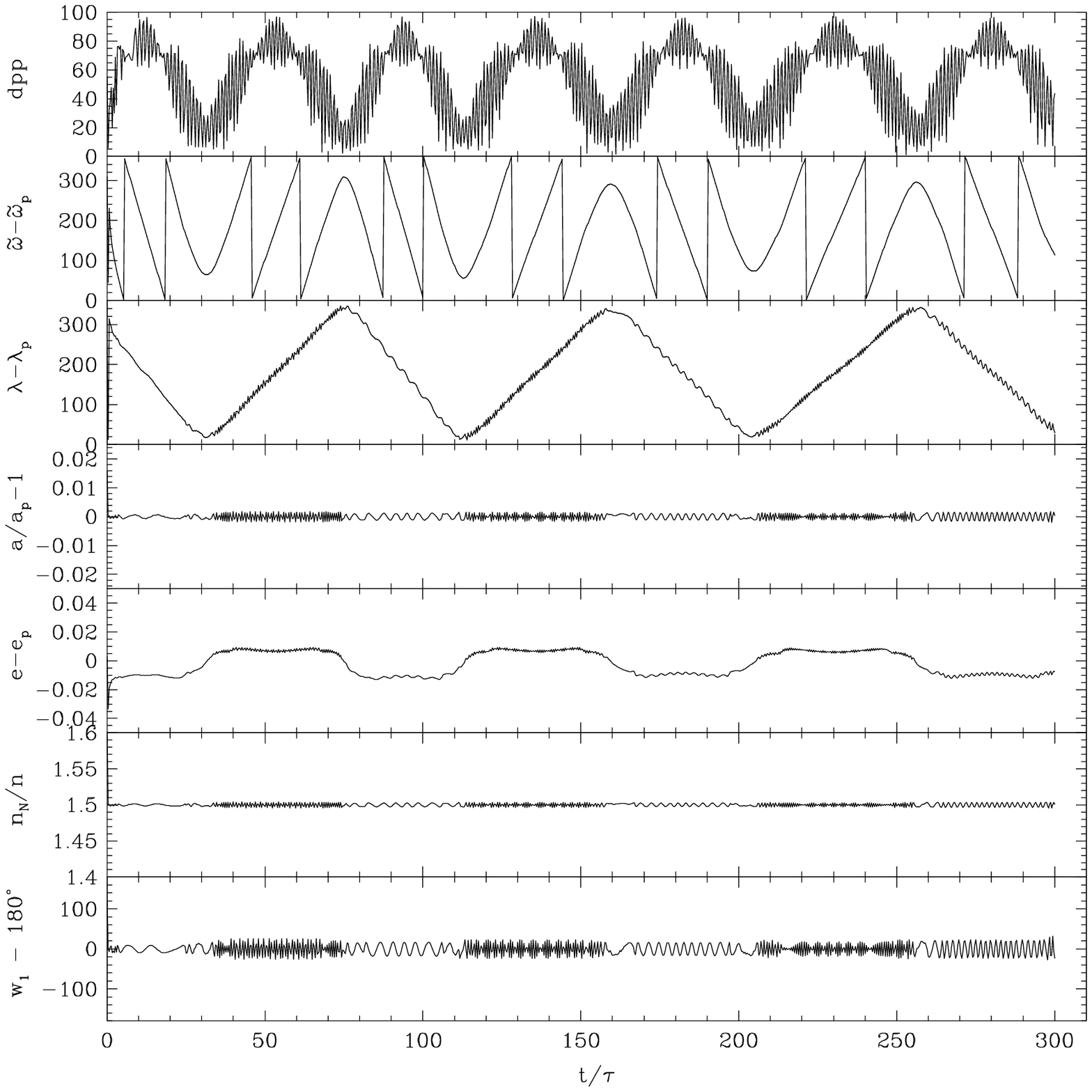]{A horseshoe orbit. \label{fig:horseshoe1}}

\figcaption[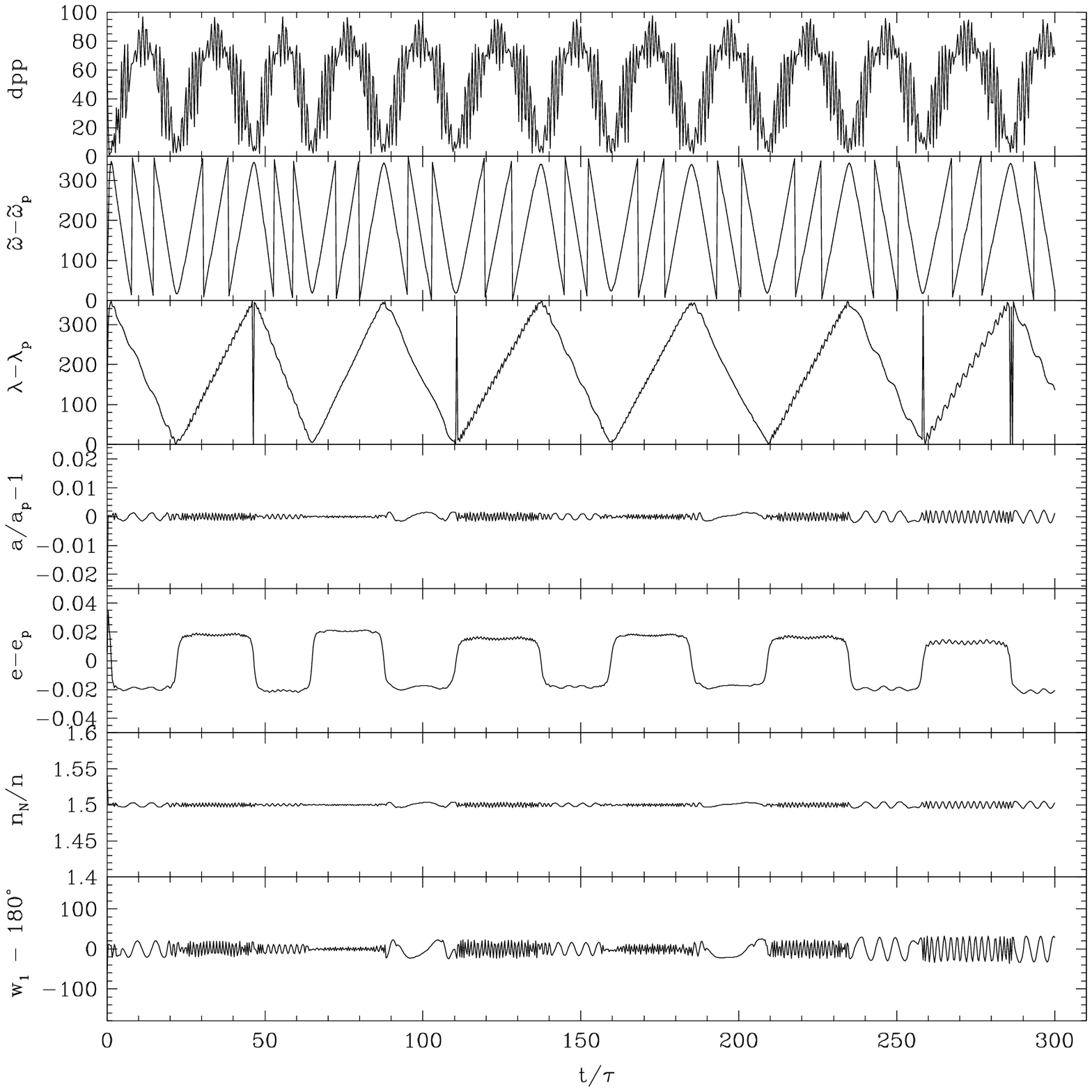]{A second horseshoe orbit. The spikes in
$\lambda-\lambda_p$ arise because small short-period oscillations occasionally
carry this angle past 0 or $2\pi$. \label{fig:horseshoe2}}

\figcaption[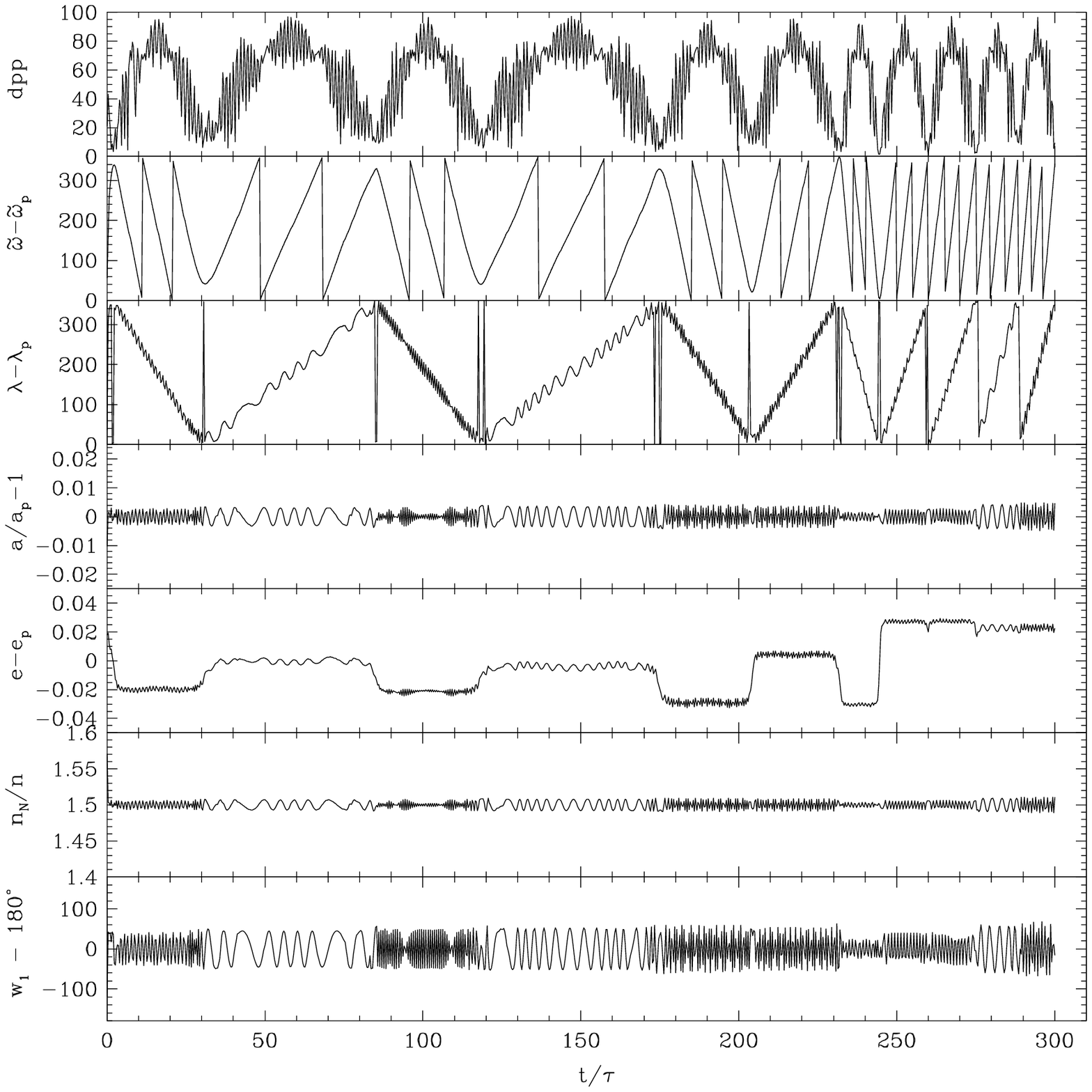]{A transitional orbit. \label{fig:trans}}

\figcaption[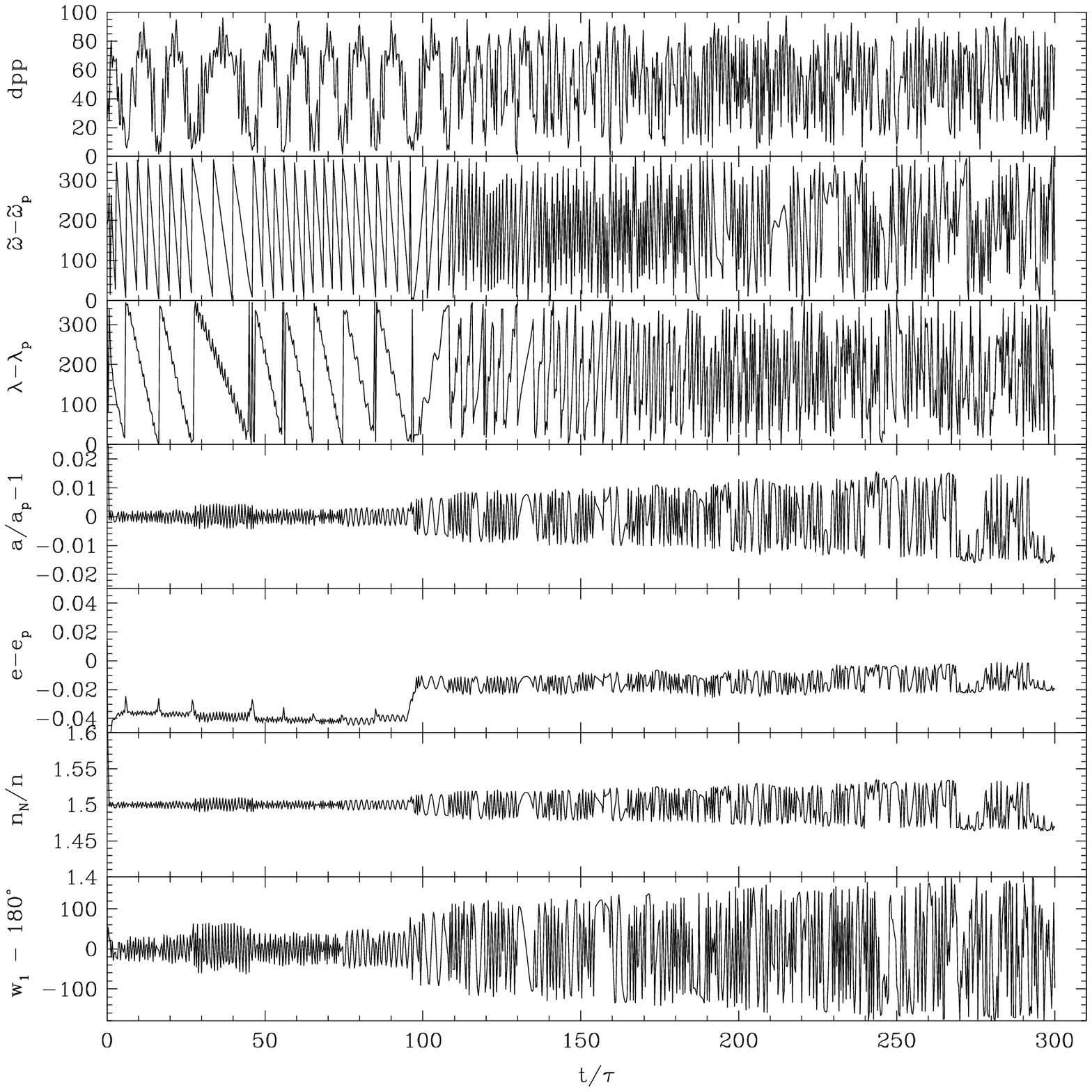]{A doubly transitional orbit. \label{fig:dtrans}}

\figcaption[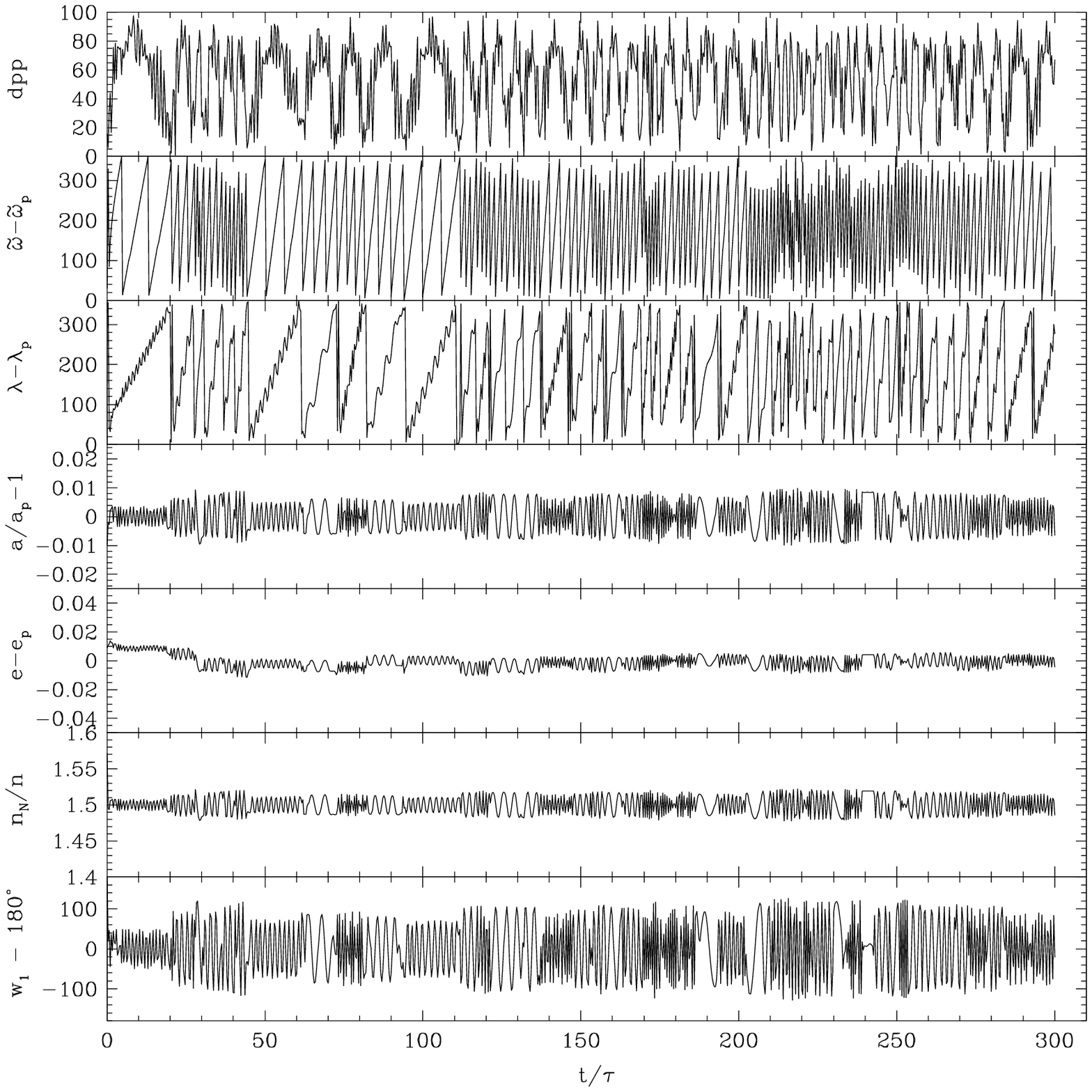]{An irregular circulating orbit. \label{fig:unaffect2}}

\figcaption[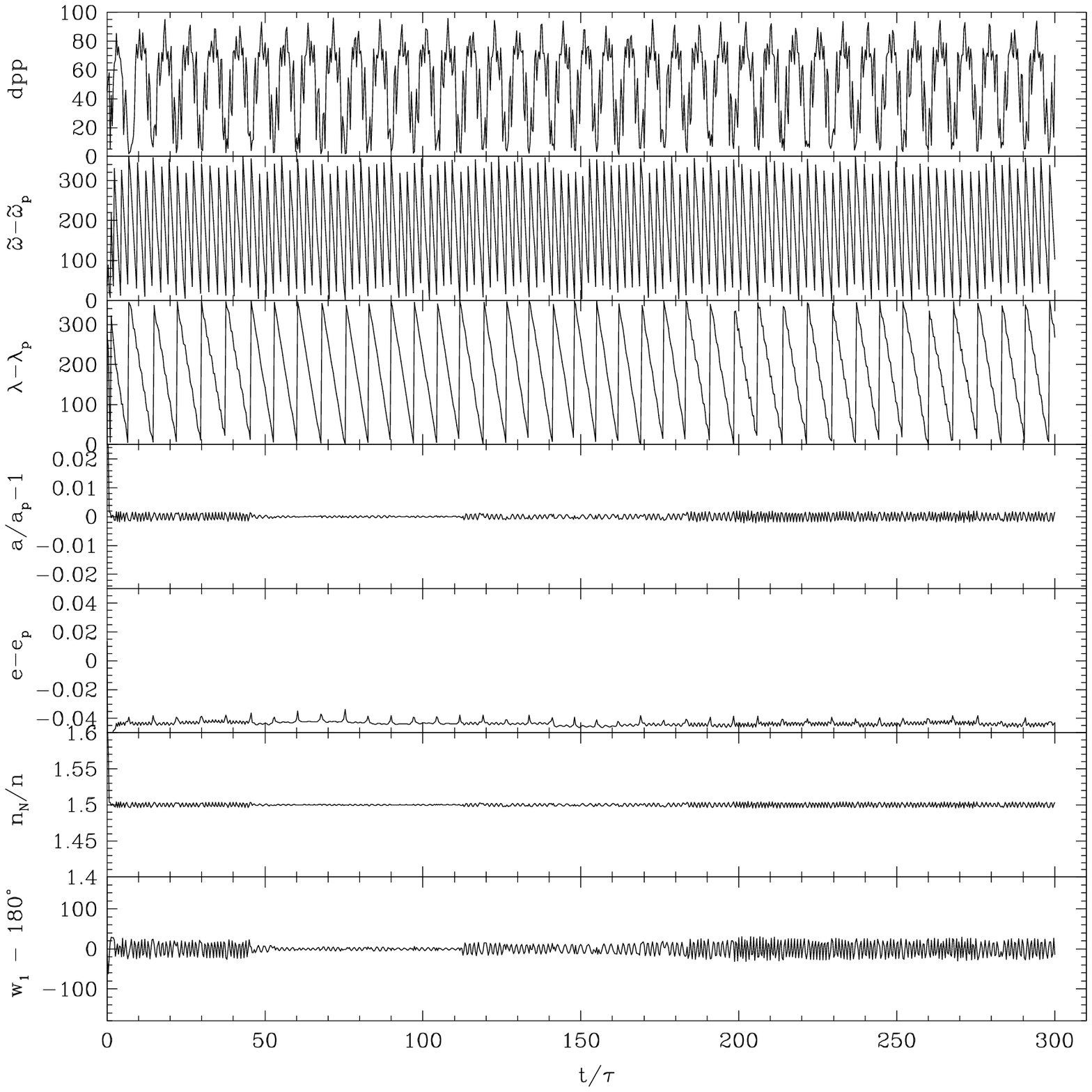]{A regular circulating orbit. \label{fig:unaffect1}}


\begin{thebibliography}{99}

\bibitem[Cohen \& Hubbard 1965]{CH65} Cohen, C. J., \& Hubbard, E. C. 1965,
AJ, 70, 10

\bibitem[Fern\'andez \& Ip 1984]{FI84} Fern\'andez, J. A., \& Ip, W.-H. 1984,
Icarus, 58, 109

\bibitem[Malhotra 1993]{M93} Malhotra, R. 1993, Nature, 365, 819

\bibitem[Malhotra 1995]{M95} Malhotra, R. 1995, AJ, 110, 420

\bibitem[Malhotra 1996]{M96} Malhotra, R. 1996, AJ, 111, 504

\bibitem[Malhotra 1998]{M98} Malhotra, R. 1998, 29th Annual Lunar and
Planetary Science Conference, Houston, TX, abstract no. 1476 

\bibitem[Malhotra \& Williams 1997]{MW97} Malhotra, R., \& Williams,
J. G. 1997, in Pluto and Charon, eds. S. A. Stern and D. J. Tholen (Tucson:
University of Arizona Press), 127

\bibitem[Malhotra et al. 1999]{MDL99} Malhotra, R., Duncan, M.,
\& Levison, H. 1999, in Protostars and Planets IV.

\bibitem[Morbidelli 1997]{mo97} Morbidelli, A. 1997, Icarus, 127, 1

\bibitem[Namouni et al. 1999]{N99} Namouni, F., Christou, A. A., \& Murray,
C. D. 1999, astro-ph/9904016

\bibitem[Stern 1992]{S92} Stern, S. A. 1992, ARAA, 30, 185

\bibitem[Tholen \& Buie 1997]{TW97} Tholen, D. J., \& Buie, M. W. 1997, in 
Pluto and Charon, eds. S. A. Stern and D. J. Tholen (Tucson:
University of Arizona Press), 193

\bibitem[Yoder et al. 1983]{Y83} Yoder, C. F., Colombo, G., Synnott, S. P., \&
Yoder, K. A. 1983, Icarus, 53, 431

\end{thebibliography}
\end{document}